\documentclass[A4]{pasj00}
\Received{$\langle$reception date$\rangle$}
\Accepted{$\langle$acception date$\rangle$}
\Published{$\langle$publication date$\rangle$} \SetRunningHead{Erlin
Qiao \&  B.F. Liu}{The Dependence of Spectral State Transition and
Disk Truncation on Viscosity Parameter $\alpha$}

\usepackage{times}
\usepackage{graphicx}
\usepackage{epsfig}
\usepackage{lscape}

\begin{document}

\title{The Dependence of Spectral State Transition and Disk Truncation on Viscosity Parameter $\alpha$ }
\author{Erlin Qiao\altaffilmark{1,2} and B.F. Liu\altaffilmark{1}}
%
\affil{%
\altaffilmark{1} National Astronomical Observatories /Yunnan Observatory,
Chinese Academy of Sciences, P.O. Box 110,
 Kunming 650011, P. R. China\\
\altaffilmark{2} Graduate School of Chinese Academy of Sciences, Beijing 100049, P. R. China}
\email{qel1982@ynao.ac.cn \\
bfliu@ynao.ac.cn} 

\maketitle

\begin{abstract}
A wealth of Galactic accreting  X-ray binaries have been observed
both in low/hard state and high/soft state. The transition between
these two states was often detected. Observation shows that the
transition luminosity between these two states is different for
different sources, ranging from 1\% to 4\% of the Eddington luminosity. Even
for the same source the transition luminosity at different outbursts
is also different. The transition can occur from 0.0069 to 0.15
Eddington luminosity. To investigate the underlying physics,
we study the influence of viscosity parameter $\alpha$
 on the transition luminosity
on the basis of the disk-corona model for black holes. We
calculate the mass evaporation rate for a wide range of viscosity
parameter, $0.1\le \alpha\le 0.9$. By fitting the numerical results, we obtain
fitting formulae for both the transition accretion rate and the
corresponding radius as a function of  $\alpha$. We find
that the transition luminosity is very sensitive to the value of
$\alpha$, $L/L_{\rm Edd}\propto\alpha^{2.34}$. 
For  $0.1\le\alpha\le 0.6$, the transition luminosity
varies by two orders of magnitude, from 0.001 to 0.2 Eddington luminosity.
Comparing with observations
 we find that the
transition luminosity can be fitted by adjusting the value of
$\alpha$, and the model determined values of $\alpha$ are mostly in
the range of observationally inferred value.
Meanwhile we investigate the truncation of the disk in the low/hard
state for some luminous sources. Our results are roughly in agreement
with the observations.
\end{abstract}

\section{Introduction}
It is well known that there are two basic spectral states, a high/soft state
and a low/hard state in accreting X-ray binaries. At the high/soft
state, the accretion is dominantly via a standard disk (Shakura \&
Sunyaev 1973); while at the low/hard state, the accretion is
dominantly via an advection dominated accretion flow (ADAF)  or a
radiative inefficient accretion flow (RIAF) (Narayan \& Yi 1994,
1995a,b; Abramowicz et al. 1995; for reviews see Narayan 2005 and Kato et al. 2008). Similar accretion models are
proposed in active galactic nuclei (Narayan \& Yi 1994, Yuan et al.
2003). Detailed fits to the observational spectra show that in the
soft/high state of black hole the thin disk extends down to the
innermost stable circular orbit (ISCO); while in the hard/low
states, the accretion flow consists of two radial zones, i.e., an
inner advection dominated accretion flow (ADAF) that extends from
ISCO (or even inside the ISCO, see Watarai \& Mineshige 2003) to some truncation radius and an outer
thin accretion disk, above which is the corona which provides hot
mass accretion flow for the inner ADAF (Esin et al. 1997).

The mechanisms to facilitate the state transition have been
investigated by Honma (1996), Manmoto \& Kato (2000) and Lu et al. (2004),
which are based on a radial conductive energy transport process, 
and by Meyer et al.
(2000b), R\`o\.za\`nska \& Czerny (2000), Spruit \& Deufel (2002),  and Dullemond \& Spruit (2005), which are based on a vertical evaporative process.  In strong ADAF principle, the transition luminosity is defined as  the critical accretion rate for an ADAF to
exist (Narayan \&Yi 1995b; Abramowicz et al. 1995) and thus is 
derived from the assumption that all of the accretion energy is 
transferred to electrons and eventually radiates away 
($q_{\rm vis}^+\sim q_{\rm ie}$). 
The disk-corona evaporation model (Meyer \&
Meyer-Hofmeister 1994; Meyer, Liu, Meyer-Hofmeister 2000a; b; Liu et
al. 2002) can naturally explain both the transition of spectral
state and truncation of the thin disk in the low/hard state. The
interaction between the disk and the corona leads to mass
evaporating from the disk to the corona. The evaporation rate
reaches a maximum value at a few hundred Schwarzschild radii. When
the accretion rate feeding the accretion in the outer disk is higher
than the maximal evaporation rate, the disk extends down to the
innermost stable orbit. When the accretion rate is lower than the
maximal evaporation rate, the disk is first truncated at the most
efficient evaporation region and then the inner disk could be
completely depleted if the accretion rate is too low. The final
steady state at high accretion rate is then the soft state dominated
by disk accretion and at low accretion rate the hard state dominated
by the ADAF/RIAF. The maximal evaporation rate corresponds to the
transition accretion rate.

Investigation of the soft-to-hard state transition for X-ray
binaries with good measurements of flux, distance and mass of the
compact object shows that the transition luminosity is at about
1-4\% of the Eddington luminosity, with a mean of 2\% (Maccarone
2003). This is in good agreement with the theoretical prediction of
2\% by the disk corona evaporation model (Meyer et al 2000a).
Nevertheless, the transition luminosity of some sources deviates
this value largely. For instance, the transition luminosity observed
for GX 339-4 reaches 0.15 $ {L_{\rm {Edd}}}$, while it is only
0.0069 ${L_{\rm Edd}}$ for GS 2000+251. Studies of
individual object, GX 339-4, show that even for the same object the
transition occurred at different luminosity during different
outbursts (Zdziarski et al. 2004, Belloni et al. 2006). Even in the same outburst, 
the transition luminosity can differ in the rise and decay phases (Miyamoto et al. 1995). If it is the
disk evaporation that causes the state transition in above cases,
there should be something changed  in the corona.

  One of the potential factors leading to different transition
luminosity is the change of viscosity parameter, which is fixed to
0.3 in previous studies (Meyer et al. 2000a;b). 
Although the effective temperature profile and thus 
 spectra do not depend on the $\alpha$ value at all in the standard disk,  
the emissivity in an optically thin corona depends 
on the density and the density is roughly inversely proportional to $\alpha$. Therefore,  the radiative 
luminosity from an optically thin corona/ADAF  depends on the value of viscosity parameter. On the other 
hand, there is a long research history of the $\alpha$ value through the study of outbursts of  dwarf novae 
and X-ray novae
(Meyer \& Meyer-Hofmeister 1983; Smak 1984; Lin et al. 1985; Mineshige \& Wheeler 1989).
 The most recent study for black holes 
(King , Pringle \& Livio 2007) yields $0.1< \alpha <0.6$ (if the viscosity is defined as $\nu={2\over 3}\alpha V_s H$). 

For the range of $0.1\le \alpha\le 0.3$, the influence of
viscosity has been studied by Meyer-Hofmeister \& Meyer (2001) for a
few number of $\alpha$ value. In
their calculations, the temperature of electrons is assumed equal to
that of ions, that is a one-temperature model. In reality, the accretion flows
around a black hole can reach a temperature of  $\sim 10^9 $K or
higher, at this situation electrons and ions decouple. Therefore, we
take the two-temperature disk corona model
 (Liu et al. 2002) and carry out more detailed calculations
 for a series of viscosity parameters in the range of
$0.1\le \alpha\le 0.9$ , which covers the recent observational
estimates of the viscosity range (King et al. 2007).

In this work, fitting formulae of the numerical results for both the
maximal mass accretion rate and the corresponding radius depending
on the viscosity parameter $\alpha$ are given. On the
basis of this, we compare our theoretical results with observations
for individual X-ray binary and AGNs. In Sect.2 we briefly describe
the physics of the corona above the thin disk and list equations. In
Sect.3 we show our numerical results
 and in Sect.4 we compare the model predictions
with observations. Our discussions and conclusion are given in
Sect.5 and Sect.6.

\section{The model}
The disk corona model for accreting black holes is established in
detail by Meyer et al. (2000a) and later this model is extended close
to the central black hole by taking into account the decoupling of
ions and electrons (Liu et al. 2002). It is further developed by
including the inflow and outflow of mass, energy and angular
momentum from and towards neighboring zones (Meyer-Hofmeister \&
Meyer 2003).  The basic physics of the model is briefly described
below.

We consider a hot corona above a geometrically thin standard disk
around a central black hole. In the corona, viscous dissipation
leads to ion heating, which is partially transferred to the
electrons by means of Coulomb collisions.
 This energy is then conducted down
into lower, cooler and denser corona. If the density in this layer
is  sufficiently high, the conductive flux is radiated away. If the
density is too low to efficiently radiate the energy, cool matter is
heated up and evaporation into the corona takes place.
 The mass evaporation goes on until an equilibrium
density is established. The gas evaporating into the corona still
retains angular momentum and will differentially rotate around the
central object. By friction the gas looses angular momentum and
drifts inward thus continuously drains mass from the corona towards
the central object. This is compensated by a steady mass evaporation
flow from the underlying disk. The process is driven by the
gravitational potential energy released by friction in the form of
heat in the corona. Therefore, mass is accreted to the central
object partially through  the corona (evaporated part) and partially
through the disk (the left part of the supplying mass). In the inner
region the evaporation becomes so efficient that all the matter
accreting through the disk is heated up and
 accreted towards the black hole via the corona.  Only when the accretion
rate is higher than the evaporation rate, can the disk survive.
Therefore, the disk is truncated at the radius where the evaporation
rate equals to the accretion rate.  In the case of accretion rate
higher than the maximal evaporation rate, the disk extends down to
the ISCO. In this work, we investigate how the maximal evaporation
rate and the corresponding radius varies with viscosity parameter
$\alpha$.
We use the model of Liu et al. (2002) and include the
modification on the energy equation.
For clarity, we list the
equations describing the physics of corona as follows:

Equation of state
\begin{equation}
\centering
 P={\Re \rho \over 2\mu}(T_i+T_e),
\end{equation}
where $\mu=0.62$ is the molecular weight assuming a standard
chemical composition ($X=0.75, Y=0.25$) for the corona.  For
convenience, we assume the number density of ion $n_i$ equals to
that of electron $n_e$, which is strictly true only for a pure
hydrogen plasma.

Equation of continuity
\begin{equation}
\centering
 {d\over dz}(\rho v_z)=\eta_M{2\over R}\rho v_R -{2z\over
R^2+z^2}\rho v_z.
\end{equation}

Equation of the $z$-component of momentum
\begin{equation}\label{e:mdot}
\rho v_z {dv_z\over dz}=-{dP\over dz}-\rho {GMz\over
(R^2+z^2)^{3/2}}.
\end{equation}

The energy equation of ions
\begin{equation}
\begin{array}{l}
{d\over dz}\left\{\rho_i v_z \left[{v^2\over 2}+{\gamma\over
\gamma-1}{P_i\over \rho_i}-{GM\over (R^2+z^2)^{1\over
2}}\right]\right\}\\
={3\over 2}\alpha P\Omega-q_{ie}\\
+{\eta_E}{2\over R}\rho_i v_R
\left[{v^2\over 2}+{\gamma\over \gamma-1}{P_i\over \rho_i}-{GM\over (R^2+z^2)^{1\over 2}}\right]\\
-{2z\over {R^2+z^2}}\left\{\rho_i v_z \left[{v^2\over 2}+{\gamma\over
\gamma-1}{P_i\over \rho_i}-{GM\over (R^2+z^2)^{1\over 2}}\right]\right\},
\end{array}
\end{equation}

where $\eta_M$ is the mass advection modification term and $\eta_E$ is the energy modification term.
We take
$\eta_M=1$ for the case without consideration of the effect of mass inflow and outflow from 
and towads neighboring zones in the corona, and
$\eta_E=\eta_M+0.5$ is a modification to previous energy equations (for details see Meyer-Hofmeister \& Meyer
2003). $q_{ie}$ is the exchange rate of energy between electrons and
ions,

\begin{equation}
{q_{ie}}={\bigg({2\over \pi}\bigg)}^{1\over 2}{3\over 2}{m_e\over
m_p}{\ln\Lambda}{\sigma_T c n_e n_i}(\kappa T_i-\kappa T_e)
{{1+{T_*}^{1\over 2}}\over {{T_*}^{3\over 2}}}
\end{equation}

with

\begin{equation}
T_*={{\kappa T_e}\over{m_e c^2}}\bigg(1+{m_e\over m_i}{T_i\over
T_e}\bigg),
\end{equation}
where $m_p$ and $m_e$ is the proton and electron mass, $\kappa$ is
Boltmann constant, $\sigma$ is the Thomas scattering cross and
$\ln\Lambda=20$ is the Coulomb logarithm.

The energy equation for both the ions and electrons
\begin{equation}
\begin{array}{l}
{\frac{d}{dz}\left\{\rho {v}_z\left[{v^2\over
2}+{\gamma\over\gamma-1}{P\over\rho}
-{GM\over\left(R^2+z^2\right)^{1/2}}\right]
 + F_c \right\}}\\
=\frac{3}{2}\alpha P{\mit\Omega}-n_{e}n_{i}L(T)\\
+\eta_E{2\over R}\rho v_R \left[{v^2\over
2}+{\gamma\over\gamma-1}{P\over\rho}
-{GM\over\left(R^2+z^2\right]^{1/2}}\right]\\
-{2z\over R^2+z^2}\left\{\rho v_z\left[{v^2\over
2}+{\gamma\over\gamma-1}{P\over\rho}-
{GM\over\left(R^2+z^2\right)^{1/2}}\right]
+F_c\right\},
\end{array}
\end{equation}

where $n_{e}n_{i}L(T)$ is the bremsstrahlung cooling rate and $F_c$
the thermal conduction (Spitzer 1962),
\begin{equation}\label{e:fc}
F_c=-\kappa_0T_e^{5/2}{dT_e\over dz}
\end{equation}
with $\kappa_0 = 10^{-6}{\rm erg\,s^{-1}cm^{-1}K^{-7/2}}$ for fully
ionized plasma. All other parameters in above equations are under
standard definition, and are in cgs units. The five differential
equations, Eqs.(2), (3), (4),
(7) and (8), which contain five variables
$P(z)$, $T_i$, $T_e$ , $F_c$, and $\dot m (z)(\equiv \rho v_z)$,
can be solved with five boundary conditions.

At the lower boundary $z_0$ (the interface of disk and corona), the
temperature of the gas should be the effective temperature of the
accretion disk.  Previous investigations (Liu, Meyer, \&
Meyer-Hofmeister 1995) show that the coronal temperature increases
from effective temperature to $10^{6.5}$K in a very thin layer and
thus the lower boundary conditions can be reasonably approximated
(Meyer et al. 2000a) as,
\begin{equation}
T_i=T_e=10^{6.5}K,\ {\rm and} \  F_c=-2.73\times 10^6 P\ {\rm at}\
z=z_0.
\end{equation}
At infinity, there is  no pressure and no heat flux. This requires
sound transition at some height $z=z_1$. We then constrain the upper
boundary as,
\begin{equation}
F_c=0\  {\rm and}
 \  v_z^2=V_s^2\equiv P/\rho={\Re\over 2\mu}(T_i+T_e)\  {\rm at}\  z=z_1.
\end{equation}
With such boundary conditions, we assume a set of lower boundary
values for $P$ and $\dot m$ to start the integration along $z$. Only
when the trial values for $P$ and $\dot m$ fulfill the upper boundary
conditions, can the presumed $P$ and $\dot m$ be taken as true
solutions of the differential equations.

Here we wish to point out that the viscosity parameter $\alpha$ we are studying in this work only appears in 
the viscous heating rate, ${3\over 2} \alpha P\Omega$,  and radial drift speed, $v_R\propto \alpha$, 
which is eventually associated with the heating and cooling rates.

\section{Numerical results}
In our calculation we fix the mass of black hole as $M= 6M_{\odot}$
and vary  the viscous parameter $\alpha$ in the range of  0.1-0.9.
For every given $\alpha$, the evaporation rate is calculated as
$\dot m_0\equiv \rho v_z$ at the lower boundary and then is
integrated in the radial one-zone region,  $\dot M_{\rm evap}\approx
2\pi R^2\dot m_0$. This integrated value represents the evaporation
rate in a given region around distance $R$ and thus can be compared
with the accretion rate.

Fig.1 shows how the evaporation rate varies with the distance from
the black hole. Here the evaporation rate is scaled by the Eddington
rate, $\dot M_{\rm Edd}=L_{\rm Edd}/{\eta c^2}= 1.39\times 10^{18} M/M_\odot\,{\rm g\,s^{-1}}$ 
(where $\eta=0.1$ is the energy conversion efficiency) and the radius is scaled by the Schwarzschild radius,
 $R_ {\rm S}=2GM/c^2=2.95
\times 10^5 M/M_\odot \, {\rm cm}$. It can be seen that for every
given $\alpha$ the evaporation rate reaches a maximum at a certain
distance and decreases very quickly  outside the maximal evaporation
region.  The occurrence of maximal evaporation rate is a result of
energy balance in the disk corona system. In the outer region, the
evaporation rate increases toward central black hole since the
released accretion energy increases ($\propto R^{-1}$), which is the
source of energy for evaporation. However, with increase of number
density in the corona, the radiation becomes more and more efficient
(since the bremsstrahlung is proportional to the square of number
density) and only a very little (or even no) gas is heated up and
evaporates. Thus, the evaporation rate reaches a maximum value and
decreases again in the inner region.

Fig.1 also shows that the value of evaporation rates increases with the value of viscosity
parameter, which is significant in the inner region but not obvious in the outer region.
The increase of evaporation rate with viscosity is understood as a
result of efficient heating and unimportant advection. In the inner region, 
the advective cooling (related to $\alpha$) is not the dominant cooling process, as shown by our numerical data. 
For larger value of $\alpha$, the
viscous heating is more efficient, which leads to more energy that
is conducted down to the chromosphere above the disk. There the
conducted energy is partially radiated away and partially used to
evaporate the cool gas. Thus, the evaporation rate enhances for
larger  $\alpha$.  This effect is not significant in  regions far
away from the black hole since most of the viscous heating is
transferred to the internal energy of corona gas, only a very small
part is conducted down and relevant to evaporation.

The maximal evaporation rate is very sensitive to the value of
$\alpha$. The values of the maximal evaporation rate
 for $\alpha$=0.1 and  $\alpha$=0.9  are  0.001 and 0.2 Eddington rate, respectively, increasing
  by a factor of about 200 from  $\alpha=0.1$ to  $\alpha=0.9$.
The location of the maximal evaporation rates for $\alpha$=0.1 and
$\alpha$=0.9 are $\sim1880R_{\rm S}$ and $25R_{\rm S}$, decreasing by a factor
of about 75 from  $\alpha=0.1$ to  $\alpha=0.9$.
 The detailed results associated with the maximal evaporation rate are listed  Table 1.

\begin{figure}
\includegraphics[width=85mm,height=70mm,angle=0.0]{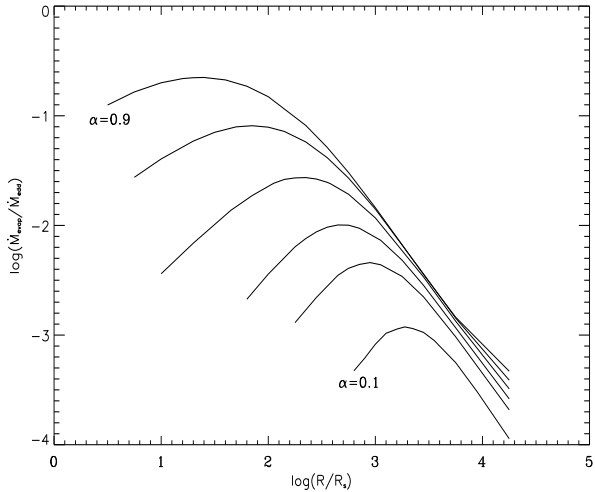}
\caption{The distribution of evaporation rate along distances for a
series of viscosity parameters. From bottom to top
$\alpha$ is 0.1, 0.15, 0.2, 0.3, 0.5, 0.9 respectively. It shows that
 the evaporation rate increases with increasing $\alpha$.  
For larger $\alpha$  the
maximal evaporation rate is higher and the corresponding radius
is smaller.}
\end{figure}

\begin{table}
\centering
\caption{The evaporation feature for different value of $\alpha$ }
\begin{tabular}{c|c|c|c}
\hline
\hline
&$\alpha$ &  $  R_{\rm max}/R_{\rm s}$
& $\dot M_{\rm max}/\dot M_{\rm Edd}$ \\
\hline
$M=6M_\odot$&0.10    &   1883.65  & $1.19 \times10^{-3}$    \\
&0.15    &   891.25  &  $4.58 \times10^{-3}$    \\
&0.20    &   446.68 &  $1.01 \times10^{-2}$ \\
&0.25    &   281.84  &  $1.78\times10^{-2}$  \\
&0.30    &   223.87 &  $2.73\times10^{-2}$   \\
&0.40    &   112.20 &  $5.16\times10^{-2}$  \\
&0.50    &   70.79 & $8.11\times10^{-2}$   \\
&0.70    &   37.58 & $1.49\times10^{-1}$   \\
&0.90    &   25.12&  $ 2.24\times10^{-1}$   \\
\hline
$M=10^8 M_\odot$ &0.3&223.87& $2.73\times10^{-2}$ \\
&0.5& 70.79&$8.11\times10^{-2}$\\
\hline
\end{tabular}\\
Note:$\dot M_{\rm max} $ and $ R_{\rm max} $ represent the maximal
evaporation rate and the corresponding radius for different
$\alpha$.\\
\end{table}

Plotting the data in Fig.2, we find that the dependence of both
the maximal evaporation rate
and the corresponding radius on the viscous parameter can be
expressed in power-law forms.
The best fits to the
data are given as
\begin{equation}\label{e:mdot-alpha}
(\dot{M}_{\rm evap}/\dot{M}_{\rm Edd})_{\rm max} \approx
0.38\alpha^{2.34},
\end{equation}
and
\begin{equation}\label{e:r-alpha}
(R / R_{\rm S})_{\dot{M} = \dot{M}_{\rm max}} \approx 18.80 \alpha^{-2.00}.
\end{equation}

\begin{figure*}
\includegraphics[width=90mm,height=70mm,angle=0.0]{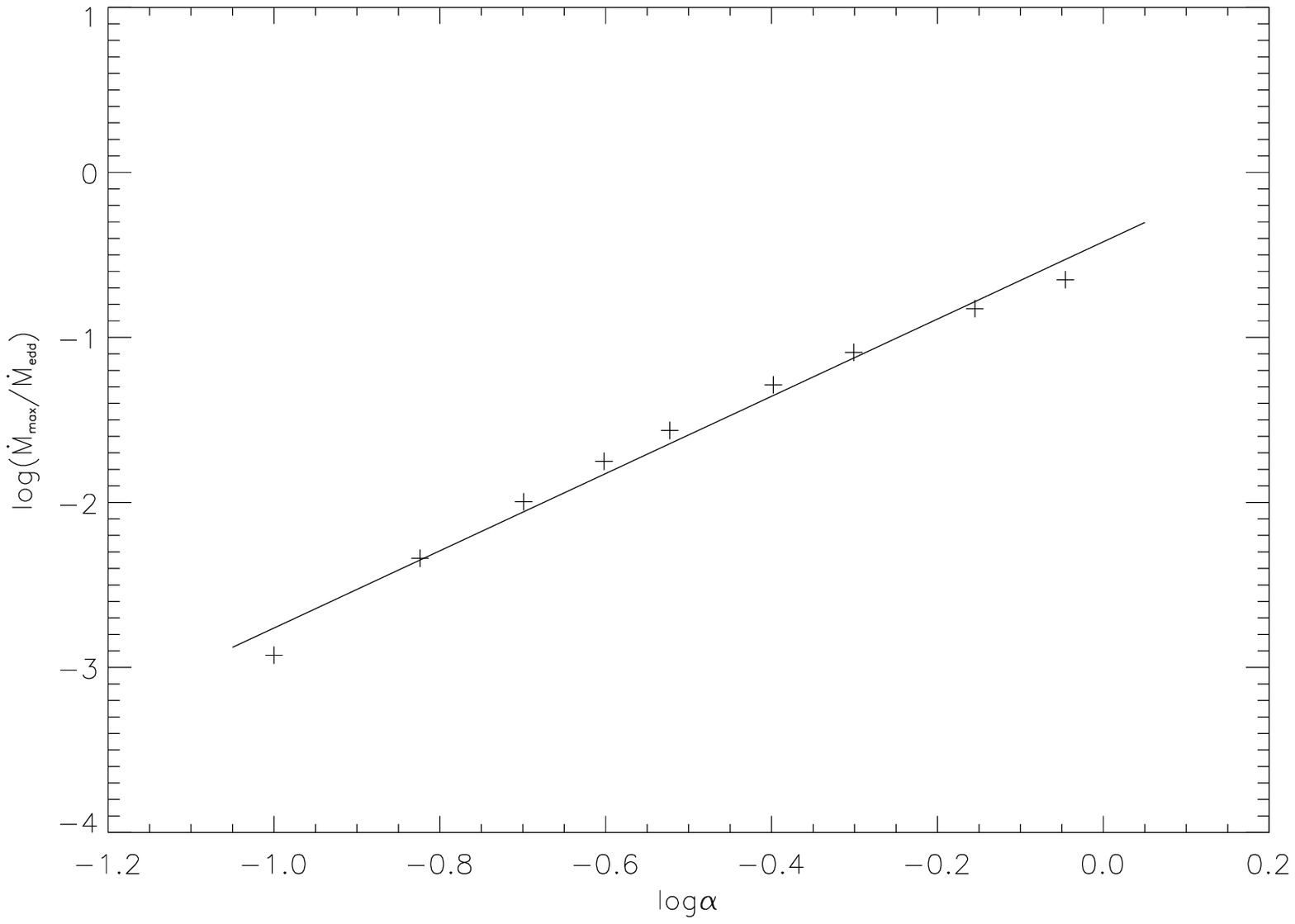}
\includegraphics[width=90mm,height=70mm,angle=0.0]{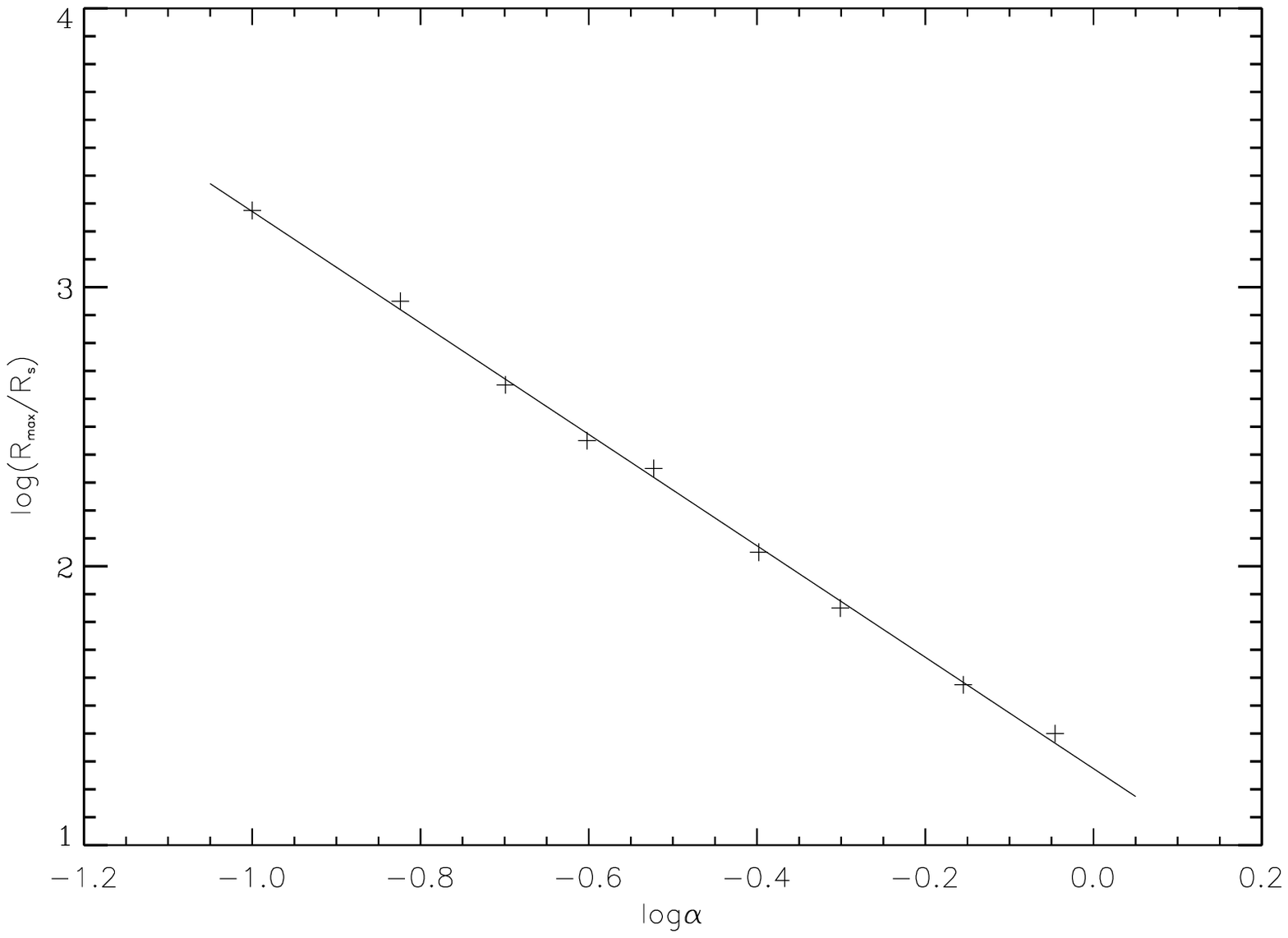}
\caption{Left:The maximal evaporation rate vs.
viscosity parameter. Right:The radius corresponding maximal
evaporation rate vs. viscosity parameter. The figure shows that the maximal
evaporation rate increases and the corresponding radius decreases with increase of viscous parameter $\alpha$}
\end{figure*}

As it has been pointed out before (Meyer et al. 2000a;b), the
existence of a maximal evaporation rate along radial direction has
very important consequence on the spectral state transition. When
the accretion rate supplying to the outer disk is lower than this
value, the disk is completely depleted first at the most efficient
evaporation region, i.e. the region where the maximal evaporation
rate reaches.  The disk is thus truncated and then recedes outwards
until the evaporation is too weak to compete with the mass flowing
rate in the disk. At the final steady state the disk is truncated at
a distance where the supplying accretion rate is equal to the
evaporation rate. In the inner region a pure corona/ADAF fills in.
Therefore, the accretion is via an inner ADAF and an outer thin
disk, and the radiation from the accretion flows forms a  hard-state
spectrum.  When the accretion rate is higher than the maximal
evaporation rate, evaporation cannot deplete the disk at any
distances, and the disk extends down to the innermost stable orbit.
Therefore, the accretion is dominated by the thin disk, and the corona
is very weak. The spectrum is dominated by multi-color blackbody.
Therefore,
 the value of maximal evaporation rate represents
 the critical mass accretion rate for the transition from
hard/low state to high/soft state. The corresponding radius
indicates how close to the black hole the thin disk is
 before the state transition.

Previous study (Meyer \& Meyer-Hofmeister 2000a, Liu et al. 2002)
reveals that the properties of evaporation are independent on the
black hole mass. To clarify this for the case of  different viscous
parameters, we also perform calculations with a black hole mass of
$10^8M_\odot$. We find that for any given $\alpha$ in the range of
$\alpha=0.1$ to 0.9, the results are mass independent as the
evaporation rate is scaled in Eddington accretion rate and the
distance in Schwarzschild radius. As an example, we list the value
of $\alpha=0.3, 0.5$, one can see from Table 1. This indicates that
our model and predictions are not only valid in stellar mass black
hole, but also are applicable to supermassive black holes like AGNs.
The most direct implication is that the disk truncation and state
transition should occur at the same accretion rate for both the
black hole X-ray binaries and AGNs if the values of $\alpha$ are
the same.

\section{Comparison with observation}
Our calculations show that for standard viscosity, $\alpha=0.3$, the
maximal evaporation rate is $\sim$2\% of the Eddington rate, which implies the transition
from hard to soft state occurs at an accretion rate $\sim$2\% of the Eddington rate. This
is in good agreement with Maccarone (2003)'s observational study on
the X-ray binaries, where it is found the average transition rate is
also $2\%$ of the Eddington rate.

The change of viscosity provides a possible explanation for the
deviation of the transition accretion rate from the average value.
For X-ray binaries, King et al. (2007) found a  range from 0.2 to
0.4, which is 0.3 to 0.6 in our definition of $\alpha$.  In this
range the disk evaporation model predicts  a range for the
transition accretion rate of 2\% to 11\%.  The outer disk can reach
down to 52$R_{\rm S}$ before transitions to the soft state. For a lower
viscous parameter the transition accretion rate can be even smaller.
For instance, in the case of $\alpha=0.2$, the predicted accretion
rate is 1\%. The strong dependence of maximal evaporation rate on
the viscous parameter indicates that  a small fluctuation of
viscosity could lead to the observational inferred  variation in the
transition rate and may explain why the spectral state transits
from hard to soft state at different accretion rates for different
objects, or for the same object at different outbursts. The
dependence of the truncation radius at state transition on the
viscous parameters also provides a possible mechanism for the change of
observational inferred disk truncation  extending down to $\sim
100R_{\rm S}$ at hard state before transition .  The numerical fitting
formulae given in this work are useful to determine the value of
the viscosity parameter from the observed transition luminosity and
then determine the truncating radius for individual objects.  On the
other hand, if viscosity can be determined by some other way,
Eqs.(\ref{e:mdot-alpha}) and (\ref{e:r-alpha}) can predict the
accretion rate at spectral transition and spectral energy
distribution contributed by the disk and the corona.

\subsection{Fits to the transition luminosity}
From observations one obtains the luminosity at the spectral
transition.  Adopting $\eta=0.1$ for the energy conversion
coefficient, we have  $ L_{\rm tran}/L_{\rm Edd}=\dot M/\dot M_{\rm
Edd}$. Thus, according to our evaporation model the transition
luminosity is a function of viscous parameter $\alpha$, that is,
\begin{equation}\label{e:L-alpha}
 L_{\rm tran}/L_{\rm Edd}\approx 0.38\alpha^{2.34}.
\end{equation}
To fit the transition luminosity, we collect data from literatures
for the sources that the transition luminosity is well measured. The
detailed results are listed in Table 2. If the variation of
transition luminosity is caused by the change of viscosity, we can
calculate the value of $\alpha$ from equation (\ref{e:L-alpha}). The
derived values are also listed in Table 2. One can see that though
the transition luminosity varies largely from about 0.0069 $L_{\rm
Edd}$ in GS 2000+251 to 0.15 $L_{\rm Edd}$ in GX 339-4, the values
of $\alpha$ are in a relatively small range from 0.18 to 0.67. Here
we must note the different definition of  $\alpha$  in different
literatures. We take $\nu= {2\over 3} \alpha V_{s}^{2}\Omega$
(Shakura \& Sunyaev 1973) for the kinematic viscosity, while  King
et al. (2007) adopt $\nu= \alpha V_{s}^{2}\Omega$ (Frank et al.
2002). Thus, the value of $\alpha$ in our work differs by a factor
of 3/2 from that of King et al. (2007). Translating the above values
of viscous parameters to the definition of King et al. (2007), the
fitted value of $\alpha$ is in the range of $\alpha$ $\sim $ 0.12 -
0.45. These values are in agreement with the observational values of
0.1- 0.4 inferred by King et al. (2007).
Here we wish to point out that there are uncertainties in the inferred transition 
luminosity $L_{\rm tran}/L_{\rm Edd}$ due to the uncertainties in measuring the mass 
of black hole and its distance. This leads to a small deviation in the value of viscosity 
parameter in fitting the transition luminosity. Nevertheless, we expect the value of $\alpha$ 
is still in the range of King et al. (2007).

With the value of viscous parameter $\alpha$, we can calculate the
truncation radius of the disk at the spectral transition from
equation (\ref{e:r-alpha}). If observations can determine the
transition radius, say, by means of emission lines, one can compare
the observationally inferred value with the model prediction,
thereby being able to check whether the transition luminosity is caused by the
change of the viscosity.

\subsection{Fits to the truncation radius in the low/hard state}
In the low/hard state of black hole accreting systems, the accretion
rate is smaller than the maximal evaporation rate, thus the disk is
truncated at a certain distance where the accretion rate is equal to
the evaporation rate.  For a standard viscosity parameter,
$\alpha=0.3$, the evaporation model predicts that the truncation radius is
larger than $224R_{\rm S}$, as listed in Table 1. This value is larger
than the values inferred from observations for some objects.
Previous investigation on the effect of magnetic fields and
conduction (Qian et al. 2007; Meyer-Hofmeister \& Meyer 2006) show
that the truncation can shift to $\sim 100R_{\rm S}$ if the magnetic field
and reduced conduction are taken into account. However, for some
objects, this truncation radius is still too large to account for the 
observation. This contradiction can be alleviated by taking into
account the fluctuation of the viscous parameter $\alpha$.
Since the truncation radius in systems with accretion rate less than $\sim $ 2-3 percent of Eddington
rate can be well explained by changing
the strength of magnetic field and thermal conduction for standard viscous parameter (Qian et al. 2007),
here we only
collect observational data with the mass accretion rate larger than
this value.

NGC 4593  is  known as a Seyfert galaxy. By fitting the continuum
spectrum of optical-UV and Fe $K_\alpha$ line profile, Lu \& Wang
(2000) find the observational evidence supporting that thin disk
truncates at $\gtrsim 30 R_{\rm s}$. The corresponding mass
accretion rate in the outer thin disk is about 0.055$ \dot M_ {\rm
Edd}$. This requires that a value of $\alpha \geq 0.44$ from
equation (\ref{e:mdot-alpha}), with which the maximal truncation
radius is about 98 $R_{\rm S}$ according to equation (\ref{e:r-alpha}).
If a large $\alpha$ is associated with strong magnetic fields, the strong magnetic fields in this object will result in a truncation radius smaller than 98 $R_{\rm S}$. 
The discrepancy of a factor of 3 between the observation and
theoretical prediction could be decreased
 by taking both the magnetic field and thermal conduction into account (Qian et al. 2007).
Therefore, we expect that the very small truncation of the thin disk
in the low/hard state of NGC 4593 is a combined effect of viscosity,
magnetic field and conduction. Note that the viscosity and conduction are mutually related to the
magnetic fields.

GX 339-4 is of well known black hole X-ray binaries. Zdziarski et al.
(2004) study the 15 outbursts of GX 339-4 from 1987 to 2004. When in
the low/hard state the highest mass accretion rate can reach $\sim $
0.26 $\dot M_{\rm Edd}$, and also at different time the accretion
rate can be  $\sim$ 0.14, 0.17, 0.07, 0.02 $\dot M_{\rm Edd}$ in the
low/hard state (Zdziarski et al. (2004). From equation
(\ref{e:mdot-alpha}) this requires the value of $\alpha $  larger
than 0.85, 0.65, 0.71, 0.49, 0.28 respectively. For these value of
$\alpha$,  the maximal truncation radius is derived from equation
(\ref{e:r-alpha}) as 26, 44, 37, 80, 231 $R_{\rm S}$. These results
are
 in agreement with the observational values of  $R_{\rm in}
\sim $20-200 $R_{\rm S}$, although the discrepancy remains.

\begin{table*}
\centering
\begin{minipage}{12cm}
\caption{Transition luminosity and fitting value of the viscosity parameter in accreting X-ray binaries}
\begin{tabular}{c@{\hspace{1.5cm}}|c|c|c}
\hline \hline
source    &   $L_{\rm {tran}}/L_{\rm {Edd}}$ &$\alpha$ &references\\
\hline
XTE J1859+226   & 0.12             &0.61  &(Gierli$\acute{\rm n}$ski \& Newton 2006)\\
XTE J1739-278   & 0.13             &0.63  &(Gierli$\acute{\rm n}$ski \& Newton 2006)\\
XTE J1650-500   & 0.02             &0.28  &(Gierli$\acute{\rm n}$ski \& Newton 2006)\\
4U 1543-47       & 0.07            &0.49  &(Gierli$\acute{\rm n}$ski \& Newton 2006)\\

Nova Mus91      & 0.031$^*$  &0.34 &(Maccarone 2003) \\
GS 2000+251      & 0.0069$^*$  &0.18&(Maccarone 2003) \\
Cyg X-1         & 0.028$^*$   &0.33 &(Maccarone 2003)\\
GRO J 1655-40   &0.0095$^*$  & 0.21  &(Maccarone 2003) \\
LMC X-3         & 0.014$^*$   & 0.24  &(Maccarone 2003)  \\

Aql X-1 & 0.019$^*$ & 0.28 &(Maccarone 2003)      \\
4U 1608-52 & 0.042$^*$ & 0.39 &(Maccarone 2003)     \\
4U 1728-34 & 0.050$^*$ & 0.42 &(Maccarone 2003)     \\
\hline
XTE J1550-564  & 0.13             &0.63 &(Gierli$\acute{\rm n}$ski \& Newton 2006) \\
XTE J1550-564  & 0.03             &0.34 &(Gierli$\acute{\rm n}$ski\& Newton 2006)\\
XTE J1550-564  & 0.034$^*$            &0.36  &(Maccarone 2003)\\

GX 339-4       &  0.15     &0.67 &(Gierli$\acute{\rm n}$ski \& Newton 2006)\\
GX 339-4       & 0.06     &0.45 &(Gierli$\acute{\rm n}$ski \& Newton 2006)\\
\hline
\end{tabular}
\\
\\
Note: (1) Aql X-1, 4U 1608-52, 4U 1728-34 are neutron
stars, the others are block holes or black hole candidates. (2) The
luminosity band in Gierli$\acute{\rm n}$ski \& Newton (2006) is
1.5-12 kev. (3)  Asterisk denotes the transition luminosity from high/soft state to low/hard state, for which the fitting value of $\alpha$ should be slightly higher if the Compton cooling by disk photons is included in our model. (4) Different transition luminosities for XTE J 1550-564  and GX 339-4
 are from different outbursts.
\end{minipage}
\end{table*}

\section{Discussion}
The disk-corona evaporation model provides a physical mechanism for
the spectral state transition between low/hard state and high/soft
state and for truncation of the thin disk in the low state.
Nevertheless, it should be pointed out that there are still
uncertainties in quantitative fits to observations. The value of the
mass evaporation rate in our model is affected by several factors,
such as the magnetic field (Qian et al. 2007), the thermal
conduction (Meyer-Hofmeister \& Meyer 2006), and the inverse Compton
scattering (Liu et al. 2002). It also depends sensitively on the
viscosity parameter $\alpha$,  as shown from our calculations.
Therefore, one should take into account all these factors in
quantitative fits to observations.

\subsection{The effect of Compton cooling}
Compton upscattering of photons  from the disk can be important in
cooling of the corona (Liu et al. (2002). At high accretion rates,
the thin disk extends down to the ISCO,  Compton cooling becomes
efficient and hence less heat is conducted downwards, which leads to
a decreased evaporation rate.  The maximal evaporation rate also
decreases, indicating that the transition luminosity from soft to
hard decreases as a result of Compton cooling. On the other hand, at
low accretion rate the disk is truncated. Compton upscattering of
disk radiation is not efficient, thereby hardly affects the
evaporation rate. Therefore, our model in this work is valid for the
transition from low/hard state to high/state. For the transition
from high/soft to low/hard state, the transition luminosity is a few
times lower than that from low/hard to high/soft transition
(Meyer-Hofmeister et al. 2005; Liu et al. 2005), producing the
observed hysteresis. It is interesting to note that different alpha values may also contribute to the
hysteresis. This makes our prediction on hysteresis closer to the observations for objects showing 
very large hysteresis (e.g. Miyamoto et al 1995). 

In the case of a weak inner disk existing at low/hard states (Liu et al. 2007), the 
Compton effect can also lead to a lower evaporation rate, but it can not change the 
maximal evaporation rate significantly.

\subsection{The combined effects of magnetic fields, conduction and viscosity}
As it has been shown in previous study (Qian et al. 2007), the
magnetic fields and conduction hardly affect the transition
luminosity, but lead to a small truncation radius at the state
transition. Here it is shown that increase of viscosity results in
not only a decreased transition radius but also an enhanced
transition luminosity. Combination of these effects can make our
model more flexible in fits to the observationally inferred
transition luminosity and and the truncation radius.
Here one should note the physical link between magnetic field strengths and viscosity 
parameter. In fact, the viscosity is mostly provided by magnetic stress, 
though we are not very clear the detailed dependence of $\alpha$ on magnetic stress.

\subsection{The value of viscous parameter $\alpha$}
Collecting and analyzing observational results from dwarf novae,
soft X-ray transients and AGNs, King et al.(2007) find the range of
viscous parameter is $0.1\le \alpha\le 0.4$. Fits to the
observational transition luminosity by our model give similar range
for the viscous parameter.  However,  the value of
$\alpha$ derived by many MHD simulations is less than 0.02. 
For instance, Stone et al. (1996) derive
$\alpha$ $<$0.01; Brandenburg et al.(1995) get 0.001 $<$ $\alpha$
$<$0.005, Hirose, Krolik \& Stone (2006) obtain $\alpha \simeq
0.016$, Hawley \&Krolik (2001) get $\alpha \sim 0.1$. The analysis
of detailed observational data  (King et al. 2007) indicates that
most MHD simulations underestimate the value of $\alpha$. 

Caution should be taken to the fact that the $\alpha$ value obtained from the 
numerical simulations depends how to calculate them, (e.g., Machida, Hayashi \& Matsumoto 2000,
Machida \& Matsumoto 2003, Machida, Nakamura \& Matsumoto 2004). 
This is because the magnetized accretion flow 
shows highly inhomogeneous structure and significant time fluctuations.  Thus, the derived alpha 
values depend how to make time and spatial averages of physical quantities. Generally speaking, 
the ratio of magnetic energy to gas energy and thus the alpha value increases with increase 
of height; the alpha values sometimes even exceed unity in corona (see, e.g., Machida et al. 2000).  
Hence, if we make the spatial average with weight of magnetic energy instead of gas density or gas energy, 
the derived alpha values tend to be higher.

\subsection{The dependence of wind loss on $\alpha$}
The fraction of mass carried away by the wind sensitively depends on the value of $\alpha$.
For $\alpha=0.3$, the fraction increases from $\sim$7\% at the maximal evaporation region ($R=10^{2.35}R_{\rm S}$) to  $\sim$9.8\% at $R=10^{4.25}R_{\rm S}$. 
For $\alpha=0.1$, the fraction increases from  $\sim$1.4\% at 
the maximal evaporation region ($R=10^{3.27}R_{\rm S}$) to $\sim$2.7\% at $R=10^{4.25}R_{\rm S}$. 
For $\alpha=0.9$,  the fraction increases from 
$\sim$16\% at $R=25R_{\rm S}$  to  19\% at  $R=10^{4.25}R_{\rm S}$.
One can see the fraction of mass carried away by the wind increases 
systemically with an increase of the value of $\alpha$. This can be understood as the higher temperature 
and pressure in the larger $\alpha$ case.

\subsection{A mechanism for the ``strong ADAF'' principle}
Our model indicates that the accretion flow changes from a
corona/ADAF to a thin disk when the accretion rate exceeds the
maximal evaporation rate. Thus, the maximal evaporation rate
represents the critical accretion rate for the accretion  taking the
form of a corona/ADAF. Our detailed calculations show that the
maximal evaporation rate depends on the viscous parameter
approximately in the form of  $\dot m\propto \alpha^{2.34}$. This
result is similar to the critical accretion rate for an ADAF to
exist, that is, $ {\dot{\rm m}_{\rm {crit}}} \propto \alpha^{2}$
(Narayan \&Yi 1995b; Abramowicz et al. 1995; Mahadevan 1997). The
similar dependent laws on the viscosity are easy to understand since
both the maximal evaporation rate and critical accretion rate for an
ADAF to exist are derived from the assumption that all of 
the viscous heating is 
 transferred to electrons and eventually radiated away 
 ($q_{\rm vis}^+\sim q_{\rm ie}$, for details see Mahadevan 1997).  
Now that the evaporation  as a consequence of disk
corona interaction can completely deplete the inner disk and turn it
into a pure corona/ADAF when the accretion rate is less than the
maximal evaporation rate, the disk evaporation model provides a
naturally physical mechanism for the ``strong ADAF'' principle, that
is, whenever the accretion rate is lower than the critical accretion
rate, the accretion is via an ADAF.

\section{Conclusion}
Using the two-temperature disk corona model, we investigate the influence of viscosity
on the evaporation feature. We find that the evaporation rate depends strongly on the value
of viscous parameter. For  $0.1\le\alpha\le 0.6$, the maximal evaporation rate varies by
two orders of magnitude, from 0.001 to 0.2 Eddington rate. The distance corresponding to
the maximal evaporation rate also varies by near two orders of magnitude. An numerical fitting formulae
of the maximal mass evaporation rate as a function of viscosity is given by fitting
the numerical data. The corresponding radius is also given as a function of $\alpha$.
Since the maximal evaporation rate represents the critical
accretion rate at the transition from low/hard to high/soft state,
we compare our model with observations.
We find that the transition luminosity can be fitted by adjusting the value of
$\alpha$, and the model determined values of $\alpha$ are mostly in
the range of observationally inferred value (King et al. 2007).
Meanwhile we investigate the truncation of the disk in the low/hard
state for some luminous sources. Our results are roughly in agreement
with the observations.
\\

We thank F. Meyer and E. Meyer-Hofmeister
for their comments.
This work is supported by the the National
Natural Science Foundation of China (grants 10533050 and 10773028).

\end{document}